\documentclass[aps,prl,english,showpacs,twocolumn]{revtex4}
\usepackage{amsmath}
\usepackage{amssymb}
\usepackage{graphics}
\usepackage{epsfig}

\newcommand\ba{\begin{eqnarray}}
\newcommand\ea{\end{eqnarray}}
\newcommand\be{\begin{equation}}
\newcommand\ee{\end{equation}}
\newcommand\nn{\nonumber}

\begin{document}

\title{Proton electron elastic scattering and the proton charge radius}

\author{G. I. Gakh}
\affiliation{\it National Science Centre "Kharkov Institute of Physics and Technology"\\ 61108 Akademicheskaya 1, Kharkov,
Ukraine }

\author{A. Dbeyssi}
\affiliation{
CNRS/IN2P3, Institut de Physique Nucl\'eaire, UMR 8608, 91405 Orsay, France} 

\author{E.~Tomasi-Gustafsson}
\email[E-mail: ]{etomasi@cea.fr}
\altaffiliation{Permanent address: {\it CEA,IRFU,SPhN, Saclay, F-91191 Gif-sur-Yvette Cedex}}
\affiliation{
CNRS/IN2P3, Institut de Physique Nucl\'eaire, UMR 8608, 91405 Orsay, France} 

\author{D. Marchand}
\affiliation{
CNRS/IN2P3, Institut de Physique Nucl\'eaire, UMR 8608, 91405 Orsay, France} 

\author{V.~V.~Bytev}
\affiliation{Joint Institute for Nuclear Research, Dubna, Russia}
\begin{abstract}

It is suggested that proton elastic scattering on atomic electrons allows a precise measurement of the proton charge radius. Very small values of transferred momenta (up to four order of magnitude smaller than the ones presently available) can be reached with high probability. 
\end{abstract}

\maketitle


The problem of the proton size has been recently object of large interest,  due to the recent experiment on muonic hydrogen by laser spectroscopy measurement of the $\nu p$(2S-2P) transition frequency  \cite{860749}. The result on the proton charge radius $r_c=0.84184(67)$ obtained in this experiment is one order of magnitude more precise but smaller by five standard deviation compared to the best value previously assumed  $r_c=0.8768(69)$ fm \cite{arXiv:0801.0028} (CODATA). Previous best measurements include techniques based on H spectroscopy, which are more precise, but compatible with electron proton elastic scattering at small values of the four momentum transfer squared $Q^2$. The most recent result from electron proton elastic scattering,  $r_c=0.879(5)_{\rm stat}(4)_{\rm syst}(2)_{\rm model}(4)_{\rm group}$ fm, can be found in Ref. \cite{arXiv:1007.5076}. 

While corrections to the laser spectroscopy experiments seem well under control in frame of QED and may be estimated with a precision better than 0.1\%, in case of $ep$ elastic scattering the best precision which has been achieved is of the order of few percent. Different sources of possible systematic errors to the muonic experiment have been discussed, however no definite explanation of this difference has been given yet (see Ref. \cite{943666} and References therein).

Recent works have been devoted to the scattering of a proton projectile on an  electron target (see Ref. \cite{arXiv:1103.2540} and references therein). The possibility to build beam polarimeters for high-energy polarized (anti)proton beams has been shown \cite{Gl97}. Experiments have been done \cite{Ra04,Oe09}, and are ongoing with the aim to understand the experimental fact that a proton beam circulating through a polarized hydrogen target gets polarized \cite{Ra93}. The possibility to polarize antiprotons beams would open a wide domain of polarization studies at the GSI facility for Antiproton and Ion Research (FAIR) \cite{FAIR,PAX}. Assuming $C$-invariance in electromagnetic interactions, the (elastic and inelastic) reactions $p+e^-$ and  $\bar p+e^+$ are strictly equivalent. 

In Ref. \cite{arXiv:1103.2540}, the cross section and the polarization observables for proton electron elastic scattering, in a relativistic approach assuming the Born approximation, was derived. The relations connecting kinematical variables in direct and inverse kinematics were given. In particular, it was shown that large polarization effects appear at beam energies around 15 GeV. Moreover the  transferred momenta are very small even when the proton energy is in the GeV range. In this work we focus on the second issue and apply to the problem of a precise and consistent determination of the proton radius. The kinematics of proton-electron scattering is extremely peculiar and interesting in this respect.

In the elastic interaction between a proton and an electron, assuming that the interaction occurs through the exchange of a virtual photon of four momentum $k=(\omega,\vec k)$, the observables can be expressed as functions of two form factors, electric $G_E$, and magnetic $G_M$, which are functions of $Q^2=-k^2$ only.

The electric form factor, $G_E(Q^2)$ in the non relativistic limit is related to the charge distribution through a Fourier transform. For small values of $Q^2$ one can develop $G_E(Q^2)$ in a Taylor series expansion: 
\be
G_E(Q^2)= 1-\displaystyle\frac{1}{6}Q^2 <r_c^2> +O(Q^2),
\ee
where one takes into account the fact that the density (being the square of the wave function) is an even function of the spatial distance $r$, whereas the scalar product $\vec k\vec r$ is an odd function.
The root mean squared radius is the derivative of the form factor at $Q^2=0$
\be
<r_c^2>=-6 \left . \frac{dG_E(Q^2)}{dQ^2}\right |_{Q^2=0}.
\ee
The value itself of $G_E(Q^2=0)$ is given by the normalization to the proton charge. 

Form factors are derived from unpolarized $ep$ scattering through the Rosenbluth separation: measurements at fixed $Q^2$ for different angles allow to extract the electric and magnetic form factors. The polarization method \cite{Re68} has been recently applied \cite{Puckett:2010ac} providing very precise measurements of the ratio $G_E/G_M$ up to large values of $Q^2\simeq 9$ GeV$^2$. The larger precision comes to the fact that in this case one measures a polarization ratio, in which radiative corrections (at first order) cancel and the systematics effect related to the beam polarization and to polarimetry are essentially reduced.

Radiative corrections and Coulomb corrections have to be applied to $ep$ scattering experiments in particular for unpolarized measurements. Besides the problems related to the fact that there is no model independent way to calculate those radiative corrections which depend on the hadron structure, and that correlations exist in extracting form factors from the Rosenbluth fit \cite{TomasiGustafsson:2006pa}, one has to face the extrapolation of the data to $Q^2=0$ as discussed in Ref. \cite{arXiv:1007.5076}. The smallest value of $Q^2$ reached in that experiment was 0.004 GeV$^2$. 

The possibility to access much smaller values of $Q^2$ is offered by the elastic reaction induced by a proton beam on an electron target. 
Let us consider the reaction
\be
p(p_1)+e(k_1)\to p(p_2)+e(k_2),
\label{eq:eq1}
\ee
where particle momenta are indicated in parentheses, and $k=k_1-k_2=p_2-p_1$.
The expression of the differential cross section for unpolarized proton-electron scattering, in the coordinate system where the electron is at rest, can be written as:
\be
\frac{d\sigma}{dQ^2}=\frac{\pi\alpha^2}{2m^2\vec p^2}\frac{\cal D}{Q^4},
\label{eq:eqSk}
\ee

\ba
{\cal D}&=&-Q^2(-Q^2+2m^2)G_M^2+2[ G_E^2+\tau G_M^2 ]
\nn\\
&&\left[-Q^2M^2+\frac{1}{1+\tau}\left(2mE-\frac{Q^2}{2}\right )^2\right ] .\label{eq:eqD2}
\ea
where $\tau =Q^2/4M^2$ and $G_{E,M}$ are the Sachs electric and magnetic form factors, $m$ is the electron mass, ${\vec p}$ is the momentum of the proton beam.

Similarly to $ep$ scattering, the differential cross section diverges as $(Q^2)^2$ when $Q^2\to 0$. This is a well known result, which is a consequence of the one photon exchange mechanism and allows to reach very large cross sections. The expression (\ref{eq:eqD2}) differs from the Rosenbluth formula \cite{Ro50}, as additional terms depending on the electron mass can not be neglected. The electric contribution to the cross section dominates, being in all the allowed $Q^2$ range $\sim 10^7$ times larger than the magnetic one.  

Let us consider the case when  $E_p=100$ MeV. The proton energy is under the pion threshold for $pp $ reactions, which helps in reducing the hadronic background. 

\begin{figure}
\mbox{\epsfxsize=8.cm\leavevmode \epsffile{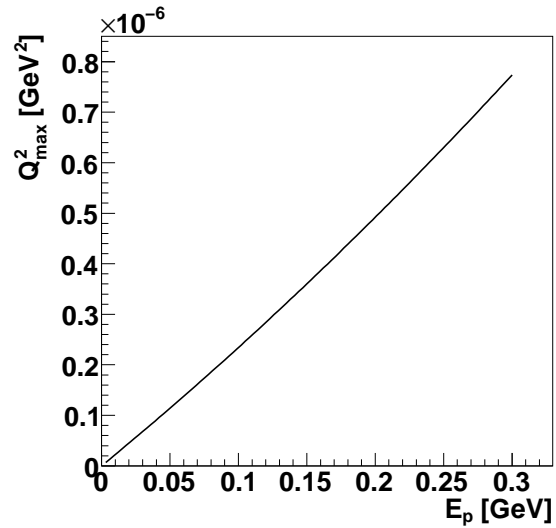}}
\vspace*{.2 truecm}
\caption{Maximum four momentum transfer squared as a function of the proton beam kinetic energy.}
\label{Fig:qmax}
\end{figure}

The properties of the inverse kinematics has been discussed in Ref. \cite{arXiv:1103.2540}. It has been shown that for a given energy of the proton beam, the maximum value of the four-momentum transfer squared is:
\be
(Q^2)_{max}=\frac{4m^2(E^2-M^2)}{M^2+2mE+m^2},
\label{eq:kmax}
\ee
where $M$ is the proton mass and $E$ is the proton beam energy. Being proportional to the electron mass squared, the four momentum transfer squared is restricted to very small values. In Fig. \ref{Fig:qmax} we report $Q^2_{max}$ as a function of the proton kinetic energy, in the MeV range. One can see that the values of transferred  momenta are very small: for a proton beam with kinetic energy $E_p=100$ MeV, $(Q^2)_{max}= 0.2\times 10^{-6}$ GeV$^2$. 

\begin{figure}
\mbox{\epsfxsize=8.cm\leavevmode \epsffile{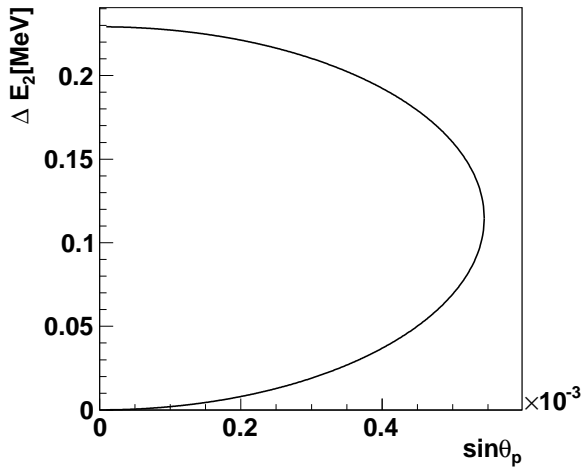}}
\vspace*{.2 truecm}
\caption{Difference of the kinetic energy of the scattered proton from the beam kinetic energy, $E_p$=100 MeV, as a function of the sine of the proton scattering angle.}
\label{Fig:Epstp}
\end{figure}

From energy and momentum conservation, one finds the following relation between the angle and the energy of the scattered electron:
\be
\cos\theta_e=\displaystyle\frac{(E+m)(\epsilon_2-m)}
{|\vec p|\sqrt{(\epsilon_2^2-m^2)}},
\label{eq:eq3b}
\ee
where $\epsilon_2$ is the energy of the scattered electron. Eq. (\ref{eq:eq3b}) shows that $\cos\theta_e\ge 0$ (the electron can never be scattered backward). In the inverse kinematics, the available kinematical region is reduced to small values of $\epsilon_2$: 
\be
\epsilon_{2,max}=m\frac{2E(E+m)+m^2-M^2}{M^2+2mE+m^2},
\label{eq:eq3c}
\ee
which is proportional to the electron mass. From momentum conservation, on can find the following relation between the kinetic energy $E_2$ and the angle $\theta_p$ of the scattered proton  (Fig. \ref{Fig:Epstp}):
\ba
E_2^{\pm}+M&=&
 [(E+m)(M^2+mE)\pm \label{eq:eqE2}\\
&&M(E^2-M^2) \cos\theta_p\sqrt{\frac{m^2}{M^2}-\sin^2\theta_p}] \nn\\
&&[{(E+m)^2- (E^2-M^2) \cos^2\theta_p}]^{-1},
\nn
\ea
which shows that for one proton angle there may be two values of the proton energy, (and two corresponding values for the recoil- electron energy and angle-, and for the transferred momentum $Q^2$). The two solutions coincide when the angle between the initial and final hadron takes its maximum value, which is determined by the ratio of the electron and scattered hadron masses, $\sin\theta_{h,max}=m/M=0.544 \cdot 10^{-3}$. Hadrons are scattered from atomic electrons at very small angles, and the larger is the hadron mass, the smaller is the available angular range for the scattered hadron. The difference between the scattered proton kinetic energy and the beam kinetic energy is shown as function of the proton scattering angle in Fig. \ref{Fig:Epstp}. The proton kinematics is very close to the beam, which makes the detection very challenging. However a magnetic system with momentum resolution of the order of $10^{-4}$ can provide at least the measurement of the energy of the scattered proton. This would allow a coincidence measurement which may help in reducing the possible background.

\begin{figure}
\mbox{\epsfxsize=8.cm\leavevmode \epsffile{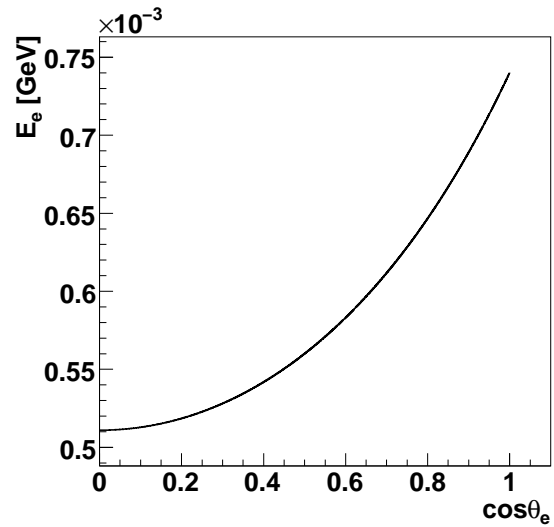}}
\vspace*{.2 truecm}
\caption{Kinetic energy of the recoil electron as a function of the cosine of the electron scattering angle for beam energy $E_p$=100 MeV.}
\label{Fig:Eece}
\end{figure}

While the proton is emitted in a narrow cone, the electron is scattered up to $90^{\circ}$. The energy dependence as function of the cosine of the angle for the recoil electron is shown in Fig. \ref{Fig:Eece}. 

In Ref. \cite{arXiv:1103.2540} it was shown that polarization observables are very small at small energy, making very difficult their measurement. Therefore, the application of the polarization method \cite{Re68} to inverse kinematics seems very challenging at low energy.  Nevertheless, the ratio of $G_E/G_M$ can be derived from the ratio of two correlation coefficients, for example $C_{tl}/C_{tt}$. Having a proton beam and an electron target both polarized in the direction normal to the scattering plane, gives access to the product of  $G_E$ and $G_M$,
once the unpolarized cross section is known:
\be
{\cal D} C_{nn}=-4mMQ^2G_EG_M.
\label{eq:eqD}
\ee

\begin{figure}
\mbox{\epsfxsize=8.cm\leavevmode \epsffile{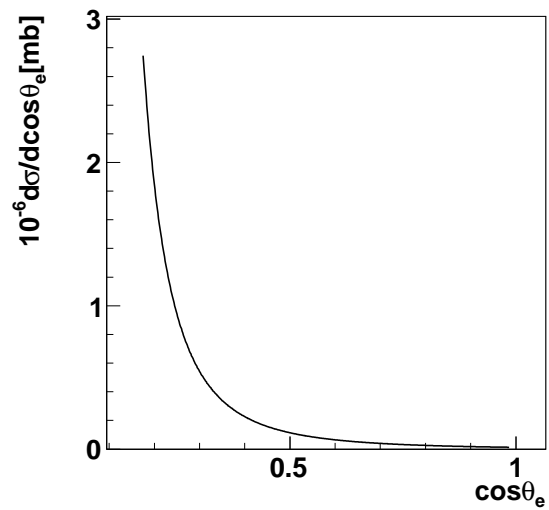}}
\vspace*{.2 truecm}
\caption{Differential cross section as a function of the cosine of the electron scattering angle for beam energy $E_p$=100 MeV.}
\label{Fig:Dece}
\end{figure}

 The differential cross section as a function of $\cos\theta_e$ is shown in Fig.  \ref{Fig:Dece} in the angular range  $10^{\circ}\le \theta_e\le 80{^\circ}$. It is large when the electron angle is close to $90^\circ$ and monotonically decreasing.
The cross section, integrated in this angular range, is $25 \times 10^4 $ mb.
Assuming a luminosity  ${\cal L}=10^{32}$ cm$^{-2}$ s$^{-1}$ with an ideal detector with  an efficiency of 100$\%$, a number of $ \simeq 25\times 10^{9} $ events can be collected in one second. Therefore, the reaction (\ref{eq:eq1}) allows  to reach very small momenta with huge cross section. The very specific kinematics, however, makes the experimental measurement very challenging. One possibility is to detect the correlation between angle and energy of the recoil electron. The detection of the energy of the scattered proton in coincidence is feasible, in principle, with a magnetic system.


In conclusions, a general characteristic of all reactions of elastic and inelastic hadron scattering by atomic electrons (which can be considered at rest) is the small value of the transfer momentum squared, even for relatively large energies of colliding hadrons. We illustrated the accessible kinematical $Q^2$ range and shown that one could improve by four order of magnitudes the lower limit at which elastic experiments have been done. In such kinematical conditions, the contribution to the cross section comes almost fully from the electric form factors. This allows a precise measurement of the proton radius, decreasing the errors due to the extrapolation to $Q^2\to 0$. However, one has to face the experimental problem of selecting elastic events, as the protons are emitted in a very narrow cone around the beam direction, with energy close to the beam one. Concrete examples of setup and realistic simulations will be object of a forthcoming paper. 

\section{Acknowledgments}
One of us (A.D.) acknowledges the Lebanese CNRS for financial support. This work was partly supported by  CNRS-IN2P3 (France) and by the National Academy of Sciences of Ukraine under PICS n. 5419 and by GDR n.3034 'Physique du Nucl\'eon' (France). L. Tassan-Got is thanked for useful discussions on experimental possibilities.


\end{document}